\documentclass[%
superscriptaddress,
twocolumn,
 amsmath,amssymb,
 aps,
prb,
]{revtex4-1}
\usepackage{ upgreek }
\usepackage{algorithm2e}
\usepackage{graphicx}
\usepackage{dcolumn}
\usepackage{bm}

\begin{document}

\preprint{APS/123-QED}

\title{Anderson localization effects on doped Hubbard model}

\author{Nathan Giovanni}
\affiliation{Departamento de F\'isica, Universidade Federal de Minas Gerais,
  C.P.~702, 30123-970, Belo Horizonte, MG, Brazil}%
\affiliation{Universit\'e Paris-Saclay, CNRS, Laboratoire de Physique des
  Solides, 91405, Orsay, France}%
\author{Marcello Civelli}%
\affiliation{Universit\'e Paris-Saclay, CNRS, Laboratoire de Physique des
  Solides, 91405, Orsay, France}%
\author{Maria C. O. Aguiar}%
\affiliation{Departamento de F\'isica, Universidade Federal de Minas Gerais,
  C.P.~702, 30123-970, Belo Horizonte, MG, Brazil}%
\affiliation{Universit\'e Paris-Saclay, CNRS, Laboratoire de Physique des
  Solides, 91405, Orsay, France}%

\date{\today}

\begin{abstract}
We derive the disorder vs. doping phase diagram of the doped
Hubbard model via Dynamical Mean Field Theory combined with
Typical Medium Theory, which allows the description
of both Mott (correlation driven) and Anderson (disorder driven)
metal-insulator transitions. We observe a transition from a metal
  to an Anderson-Mott insulator for increasing disorder strength at all interactions.
  In the weak correlation regime and rather small doping, the Anderson-Mott insulator
  displays properties which are alike to the ones found at half-filling.
  In particular, this phase is characterized by the presence of empty sites.  If we further 
  increase either the doping or the correlation however,
  an Anderson-Mott phase of different kind arises for sharply weaker disorder strength.
  This phase occupies the largest part of the phase
  diagram in the strong correlation regime, and
  is characterized by the absence of the empty sites.
\end{abstract}

\maketitle


\section{\label{sec:intro}Introduction\protect\\}

Mott proposed that electronic correlations can drive a system
through a metal-insulator transition (MIT).\cite{mott1974metal} Hubbard showed
that in a half-filled lattice this transition happens when the local
correlation contribution is larger than a critical
value.\cite{hubbard1964electron} Large correlations are present when a
material has a narrow valence band, in which case electrons spread less
in the lattice and thus interact more between them, favoring the formation
of a Mott insulator.\cite{imada1998metal,dagotto2005complexity} Transition
metal oxides are examples of materials where this Mott physics plays
a key-role.\cite{dagotto2005complexity,mcwhan1973metal,limelette2003universality}

In the opposite limit 
i.e. non-interacting electrons, the presence of disorder
can also drive the systems into an insulating phase - the Anderson insulator
in this case.\cite{anderson1958absence,lee1985disordered}
Even though there have been improvements in sample growing techniques,
effects of disorder are hardly avoidable.
Therefore, in doped Mott systems too, disorder plays a non-trivial role,
interplaying with doping 
and correlation. These effects are
hard to analyze from both experimental
and theoretical perspectives.

In experiments, correlation and disorder effects interplay, for example,
in the MIT observed in doped semiconductors, such as Si:P and
Si:B,\cite{vladeduardoreview} and in dilute two dimensional electron and
hole systems, like silicon metal-oxide-semiconductor field-effect
transistors (MOSFET’s) and semiconductor
heterostructures.\cite{elihu2001,sarachikbook} More recently, the
observation of disorder induced insulator to metal transition has been
reported in Mott systems, such as layered dichalcogenide 1T-TaS$_2$\cite{nandiniTa}
and Ru-substituted Sr$_3$Ir$_2$O$_7$.\cite{wang2018disorder}

From the theoretical viewpoint, the interplay between correlation
and disorder can be well described 
by the Hubbard model solved within extensions of Dynamical Mean Field
Theory~(DMFT).\cite{RMP_DMFT_1996} DMFT description of disorder is
equivalent to that of the coherent potential approximation~(CPA)\cite{ziman}
and, as such, misses to describe Anderson localization
effects.\cite{anderson1958absence} To circumvent this problem,
a mean field treatment of disorder, the so-called
Typical Medium Theory~(TMT),  has
been proposed and proved capable of describing the disorder-induced
localization.\cite{TMT,TMT2015,tmt_dft} The combination of TMT with DMFT
has contributed to our
understanding of the non-trivial interplay between correlation and disorder
localization effects.\cite{byczuk2005mott,Carol2,ByczukAFM,helenatmt,ByczukSpin,ekuma_clusterTMT,jarrell2016,jarrell2018}

In previous works based on DMFT-TMT, an insulating phase which is a
mixture of Mott and Anderson insulators has been observed at
half-filling.\cite{Carol2,Carol_twofluids} This Anderson-Mott insulator
is characterized by the presence of singly-occupied sites, like in a Mott insulator,
but has also doubly-occupied and empty sites, like in an Anderson
insulator.
We shall hereby refer to this Anderson-Mott insulator as AMI-0.
Here, we extend these works by investigating the doped-dependent phase diagram of the
disordered Hubbard model.

According to our results, in the small correlated regime and moderately low doping, disorder induces an
AMI-0 which is alike to the one found in the disorder-driven Anderson-Mott transition at half-filling.
As the number of carriers and/or the electronic correlation
increases however, the empty sites become occupied and an Anderson-Mott insulator (AMI) different from
the AMI-0 observed at half-filling arises. In the strongly correlated regime,
the AMI sets in a large part of the disorder versus doping phase diagram at much
weaker disorder strengths than the AMI-0. This shows in particular that the
doped strongly correlated metal is more susceptible to Anderson-Mott
induced localization than the weakly correlated metal.

The paper is organized as follows. In the next section we define the
model and describe the methodology used to solve it. Section \ref{sec:PhaseDiagram} is
devoted to the presentation and discussion of the disorder versus
doping phase diagrams built for different values of the electronic
correlation. In subsection~\ref{Vshaped}, we study the AMI-0 region of the phase diagrams,
that displays properties similar to the ones of the AMI-0 known at half-filling.
In Sec. \ref{sec:results} we explore in details the results
obtained in the strong correlation regime $U/4t=3$
and characterize the rising of a different disorder-driven AMI,
which has no empty sites. Finally,
Sec.~\ref{sec:conclusion} contains a summary of our conclusions.

\begin{figure*}[thb!!]
\includegraphics[scale=0.64]{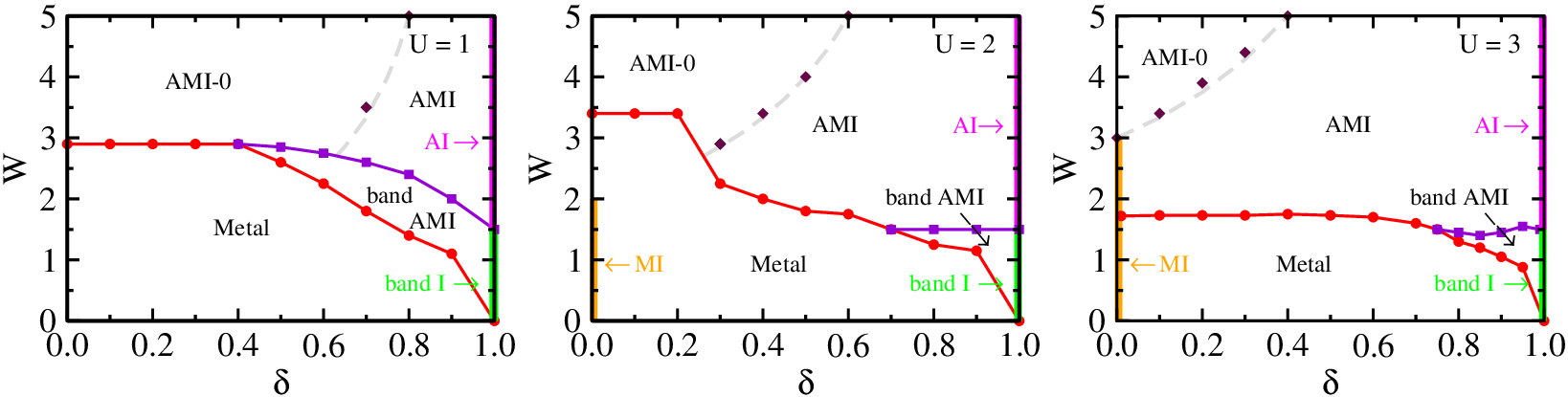}
\caption{\label{fig:WxdopingHU} Disorder ($W$) versus doping ($\delta$) phase diagram 
  of the doped Anderson-Hubbard model obtained within DMFT-TMT for $U=1$, $U=2$, and $U=3$
  at $T=0.01$. ``A" stands for Anderson, ``M" for Mott, and ``I" for insulator. Inside the
  Anderson-Mott insulator phase, we have a region where empty sites are present 
  in the system, identified as AMI-0 in the figure, and
  another region where empty sites are absent, referred to as AMI. A third
  region inside the Anderson-Mott insulator is that of a band AMI. See text
  for a complete description of the different phases and regions.}
\end{figure*}

\section{\label{sec:model}Model and methodology}

We focus on the effects of doping the Anderson-Hubbard 
model (AHM), which is given by the Hamiltonian
\begin{eqnarray}
  H&=& -t \sum_{\langle ij \rangle \sigma} (c_{i \sigma}^{\dagger}c_{j \sigma} +
  c_{j \sigma}^{\dagger}c_{i \sigma}) \nonumber \\ 
   &+& U \sum_i n_{i \uparrow} n_{i \downarrow} + \sum_{i \sigma} (\varepsilon_i -\mu) n_{i \sigma},  
\label{hamiltonian}
\end{eqnarray}
where $c^{\dagger}_{i \sigma}$ ($c_{i \sigma}$) creates (destroys) an electron with spin 
$\sigma$ on site $i$, $n_{i \sigma}= c^{\dagger}_{i \sigma} c_{i \sigma}$, 
$t$ is the hopping amplitude for nearest neighbor sites, $U$ is the on-site repulsion,
$\varepsilon_i$ is the random on-site energy, which follows a uniform probability distribution 
$P(\varepsilon)$ centered in $\varepsilon=0$ and of width $W$.
$\mu$ is the chemical potential,
which sets the doping according to $\delta=2 \left< n_{i \sigma} \right>_{av}-1$,  with respect to
the parent compound ($\delta=0$) which is nominally at half-filling
$\left< n_{i \sigma} \right>_{av}=1/2$;
  here $\left< .. \right>_{av}$ denotes an arithmetic average over disorder of the
  expectation value of the occupation number operator.
We fix here and through the paper 
  the non-interacting bandwidth $4t=1$ as energy unit.
Temperature is set to  $T=0.01$.
We consider the paramagnetic solution of the model, observed experimentally, for
example, in V$_2$O$_3$ at high temperatures.\cite{V2O3_1969,V2O3_1971}

To be able to describe the correlated Mott transition, we use DMFT.\cite{RMP_DMFT_1996}
In this methodology, a clean lattice
problem is mapped onto an auxiliary single-impurity problem, whose
conduction electron bath is determined self-consistently. In the
disordered case, the mapping is onto an ensemble of impurity problems,
each corresponding to a different value of the parameter that is randomly
distributed [on-site energy $\varepsilon_i$ in eq.~(\ref{hamiltonian})].
DMFT self-consistency involves taking an (arithmetic) average
over this ensemble. However, average values do not describe well the
asymmetric distributions generated by strong disorder.\cite{anderson1958absence}
As a drawback,
DMFT is not able to capture the Anderson transition. By considering 
the most probable or typical value over the ensemble, instead of the
average one, TMT treatment of disorder has been proved capable of describing
Anderson localization.\cite{TMT,TMT2015} Here we use the
combination of DMFT and TMT to solve the AHM (eq.~\ref{hamiltonian}) and describe
the interplay between correlation and disorder induced localization.

Within DMFT-TMT, all the impurities of the ensemble ``see'' a typical
effective medium, which is self-consistently calculated, as follows. 
We start by considering an initial function $\Delta(\omega)$ that 
describes this effective medium. By solving the ensemble 
of single-impurity problems in the presence of this bath, we obtain the 
self-energies $\Sigma_i(\omega)$ and the local Green's functions
\begin{equation}\label{Gwe}
G(\omega,\varepsilon_i) = [\omega + \mu - \varepsilon_i - \Delta(\omega) - \Sigma_i(\omega)]^{-1},
\end{equation}
from which local spectra
$\rho(\omega, \varepsilon_i)=- \frac{1}{\pi} \mbox{Im}G(\omega, \varepsilon_i)$
are calculated.
In each DMFT-TMT iteration, an effective medium is calculated: this is
given by the {\it typical or most probable value} of local impurity spectrum,
estimated by taking a {\it geometric average} over the different impurity problems. 
Precisely, the typical DOS is obtained by the geometric average
of $\rho(\omega, \varepsilon)$,
\begin{equation}
\rho_{typ}(\omega)=exp\left[\int d\varepsilon P(\varepsilon) \mbox{ln} 
\rho(\omega, \varepsilon)\right].
\end{equation}
The typical Green's function is then calculated through a Hilbert transform,
\begin{equation}
G_{typ}(\omega)= \int_{-\infty}^{\infty} d \omega' \frac{\rho_{typ}(\omega')}{\omega- \omega'}.
\end{equation}
As reference case, we consider the Bethe lattice with infinite coordination number,
which corresponds to a semicircular DOS in the non-interacting limit.\cite{RMP_DMFT_1996} 
In this particular case, we close the self-consistent loop by obtaining the new
  bath function as $\Delta(\omega)=t^2G_{typ}(\omega)$.


The typical DOS $\rho_{typ}(\omega)$ takes into account only extended states of
the system. It is thus critical at the disorder-induced transition, as the
system states become localized and $\rho_{typ}(\omega)$ is then expected to go to
  zero. The (arithmetic) average DOS, which is for instance directly detected in
  spectroscopic experiments, can also be
calculated from the DOS of the single-impurity problems, as follows:
\begin{equation}\label{eq:rhoAve}
    \rho_{av}(\omega)=\int d\varepsilon P(\varepsilon)\rho(\omega,\varepsilon).
\end{equation}
It considers both extended and localized states of the system\cite{Carol2}
and remains finite at disordered induced MIT. This quantity however goes to zero
around the Fermi level by increasing correlation.
This signals the correlated-driven Mott MIT.\cite{RMP_DMFT_1996}

Since the DMFT-TMT self-consistent condition is based on the DOS,
it is advantageous to solve the single-impurity problems on the real axis, 
to avoid an analytic continuation procedure. Here,
we solve these auxiliary
problems by using perturbation theory in $U$.\cite{Kajueter,Potthoff}
Though this method is an approximate solution of the impurity problems
  which has some well-known drawbacks (e.g. it misses the value of the
  Kondo temperature), it provides a correct description of the correlated
  metal to insulator transition, which is the goal of this study.
Away from half-filling, the modified second order perturbative
contribution in $U$ is given by an expression that interpolates between the known
results at high frequencies and at the atomic limit.\cite{Kajueter,Potthoff}
Comparisons of this approximation with exact diagonalization\cite{Potthoff}
and QMC\cite{Nolting} results give us confidence in it.
Besides directly providing the spectra, this method has the crucial advantage of
being numerically fast to allow us to build the phase diagram of
disordered problems. For each set of the model parameters,
we typically solve hundreds of single-impurity problems in each DMFT-TMT
iterative step. The single-impurity code used in this work was developed
by Jaksa Vu\v{c}i\v{c}evi\'c and Darko Tanaskovi\'c, from
the Institute of Physics in 
Belgrade, Serbia, and was previously used by one of us in
Ref.~\onlinecite{helenatmt}.

When entering into the AMI and AMI-0 regions of the phase diagram,
$\rho_{typ}(\omega)$ goes to zero,
  i.e. $\Delta(\omega)\to 0$, which means that effectively the
  Green's function in eq.~(\ref{Gwe}) reduces to the atomic limit one:\cite{hewson}
  \begin{equation}
    G_{d\sigma}(\omega)=\frac{1-\left< n_{i -\sigma} \right>}{\omega-\varepsilon_d}
    +\frac{\left< n_{i -\sigma} \right>}{\omega-\varepsilon_d-U}, \label{atomic}
  \end{equation}
  where $\varepsilon_d=\varepsilon_i - \mu$ is the impurity local energy
  (see eq.~\ref{hamiltonian}),
  and
  $\left< n_{i \sigma} \right>$ depends on the Fermi-Dirac distribution: 
  \begin{equation} \label{nd}
    \left< n_{i \sigma} \right>=\frac{1/2}{1+e^{(\varepsilon_d+U)/T}}+
    \frac{1/2}{1+e^{\varepsilon_d/T}}.
  \end{equation}
  $\left< n_{i \sigma} \right>$ can describe singly-occupied sites
    (typical of a Mott insulator) as well as doubly-occupied
  and empty sites (typical of an Anderson
  insulator); these are the occupations that can appear
  in the AMI and AMI-0 regions, according to the system parameters ($U$, $W$, $\delta$),
    as we mentioned in the Introduction.
  We shall plug the explicit analytical expressions (\ref{atomic})
    and (\ref{nd}) into eq. (\ref{eq:rhoAve}) to
    calculate directly $\rho_{av}(\omega)$ in the AMI and AMI-0 regions.
    This prevents eventual spurious oscillations that could appear
    from the discrete finite number of impurities considered,
    and which can spoil the correct interpretation of $\rho_{av}(\omega)$, especially close
    to the Fermi level.
In the next sections,
we shall analyze both quantities $\rho_{typ}(\omega)$ and $\rho_{av}(\omega)$ and characterize
the phases appearing in the AHM phase diagram.


\section{Disorder vs. doping phase diagrams} \label{sec:PhaseDiagram}

In Fig.~\ref{fig:WxdopingHU}, we present the disorder $W$ vs. doping $\delta$ phase diagram
  of the doped AHM obtained for three different values of correlation: weak correlation $U = 1$,
intermediate correlation $U = 2$, and strong correlation $U = 3$.

For rather small disorder ($W<3$), at half-filling the three cases analyzed in Fig.~\ref{fig:WxdopingHU} are
different:\cite{helenatmt} for $U=1$ the system is in a metallic phase, for
$U=3$ it is in a Mott insulating phase (represented by the orange line at $\delta=0$
in the phase diagram), and $U=2$ is an intermediate case, since it is a Mott
insulator for small disorder (orange line in Fig.~\ref{fig:WxdopingHU}) and
a metal for intermediate $W$. It is well known that
in the correlated case, upon doping the Mott insulator,
states appear at the Fermi level.\cite{RMP_DMFT_1996}
Thus, for all the three values of $U$ in Fig.~\ref{fig:WxdopingHU} we observe a
correlated metallic phase for a large range of doping and small disorder.

As disorder increases, a transition to Anderson-Mott insulator
takes place in the three cases at a critical disorder $W_c$.
For $U=1$ and $U=3$ {\it at small and
intermediate values of doping}, 
$W_c$ 
is practically doping independent.
We remark that $W_c$ is smaller for $U=3$ than for $U=1$,
  and in particular for the former case $W_c$ 
is smaller than $W_c \simeq U$, the critical disorder
value separating the correlated Mott insulator from the AMI-0
at half-filling (end of the orange line
  in the phase diagram).
If we now look at the results for $U=2$, the (red)
$W_c$ line 
shows a dependence with doping:
it is close to the half-filling value at small doping and, by increasing $\delta$, it decreases towards
the same value observed for $U=3$  (compare $W_c$ for $U=2$ and $U=3$ at
$\delta \approx 0.7$). $W_c$ vs. $\delta$ for $U=2$ thus
interpolates between what is observed for $U=1$ and $U=3$.

\begin{figure}
  \includegraphics[scale=0.75]{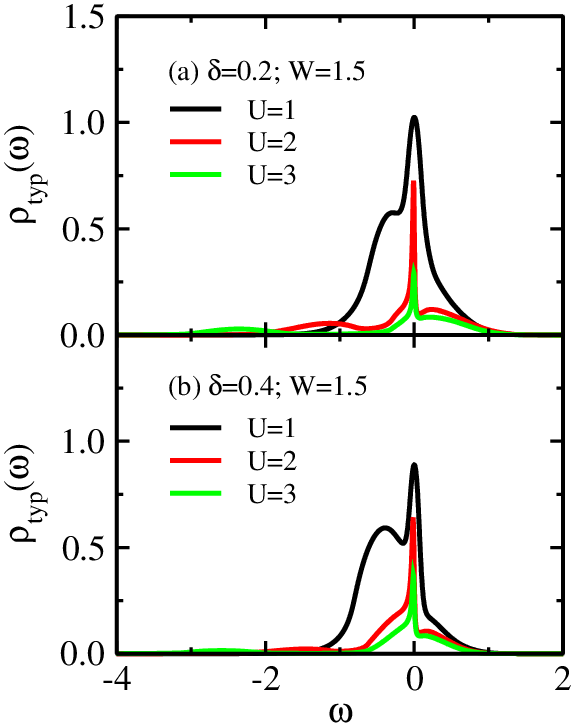}
  \caption{Comparison of the typical DOS as a function of frequency for $U=1$,
  $U=2$, and $U=3$ at a disorder value chosen such that the systems are in
    the metallic phase, $\delta=0.2$ in panel (a) and $\delta=0.4$ in panel (b).
    Although disorder is $W=1.5$ in all cases, as
  $U$ increases the system is closer to the MIT.
 } \label{compU1U3}
\end{figure}

The comparison between the weak ($U=1$) and strong ($U=3$) correlated phase
diagrams described above seems then to suggest that the metal that appears
upon doping the Mott insulator is more susceptible to disorder induced
localization than the metal which sets at small $U$.
At larger $U$, correlations strongly reduce the quasiparticle bandwith at the Fermi level
  (by a factor of $Z$, the quasiparticle
  residue), by transferring spectral
  weight from low to high energies. This forms the well known Mott peak-Hubbard
  band electronic structure.\cite{RMP_DMFT_1996} In this region of the phase diagram, 
  the narrow peak indicates that quasiparticle are less itinerant, therefore
  more easily localized by disorder. A critical $W_c$, smaller than that in the weak
  correlation region, is sufficient to localize quasiparticles.
Indeed, if we compare the typical DOS for the different values of $U$
at $W=1.5$, shown in Fig.~\ref{compU1U3}, we observe that $\rho_{typ}(\omega)$
decreases as $U$ increases. The transition occurs when
$\rho_{typ}(\omega)$ vanishes, since this indicates that extended
states of the system fully localize (we will discuss this in more detail in
Sec. \ref{mitU3}). This implies that the system with $U=3$
is closer to Anderson MIT than the one with $U=1$.

The comparison between $U=1$ and $U=3$ results indicate that the metal-Anderson-Mott insulator
  transition in the weak
  correlation, where the AMI-0 arises (see next subsection for details),
  and the transition in the strong correlation regime, where the 
  AMI arises, are sharply different.
  The AMI in fact is the byproduct of the interplay of disorder with correlation and doping,
    which suppress empty sites.   
  We shall discuss this in more detail in Sec. \ref{sec:results}.
For even larger $U$, the properties of the system are similar to the
ones at $U=3$ when we consider the same doping and disorder values (see Appendix \ref{largeU} for details).


Moving now to large doping, for each $U$ value it appears a region that
we identify as a {\it band} AMI. In this case (see section \ref{mitU3} for further details),
while electronic states are always present at the Fermi level ($\rho_{av}(\omega=0)\neq 0$), the
  typical DOS that describes the extended states shifts
  to negative frequency and acquires a zero value
  at the Fermi level, $\rho_{typ}(\omega=0)= 0$, like the DOS of a band insulator
  typically does. 

By further increasing disorder $W$, at the strong correlation regime and small doping,
the system crossovers from the AMI to the AMI-0, where empty sites are present. This AMI-0 appears
at all values of the interaction $U$ (as displayed in Fig. \ref{fig:WxdopingHU}),
  at an Anderson-Mott-like MIT in the small correlation 
  regime ($U=1$) and as a crossover within the Anderson-Mott insulating phase in the
  strong correlation regime ($U=3$). We discuss the properties of this region in detail
  in the following subsection.


\begin{figure}
\includegraphics[scale=0.75]{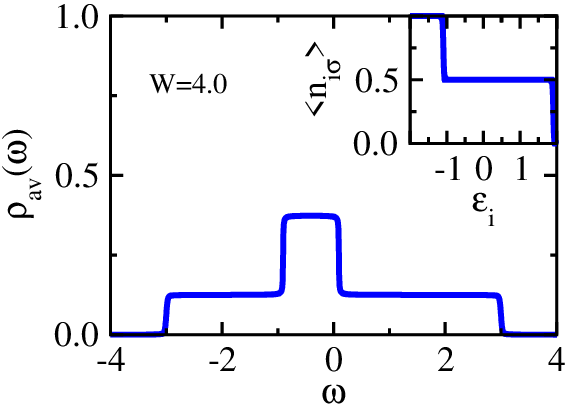}
\caption{\label{fig:VAMI_1}
  Average DOS $\rho_{av}(\omega)$ as a function 
of frequency for $W=4$ and $\delta=0.2$, that is, for parameters inside the AMI-0 region of the phase diagram. Inset: occupation number per site and per spin as a function of the on-site energy. $U=3$ and $T=0.01$.}
\end{figure}

\subsection{The AMI-0 region} \label{Vshaped}

We explore in this subsection how the AMI-0 arises within the Anderson-Mott insulating
phase, in the large disorder region of the $U=3$ phase diagram of Fig.~\ref{fig:WxdopingHU}.
This phase is the same kind of insulator that arises at the Anderson-Mott localization
at half-filling (see the results for $W=3.5$ in Appendix \ref{half-filling}), as we shall show by displaying the spectral function.

In Fig.~\ref{fig:VAMI_1} we show $\rho_{av}(\omega)$ as a function
of frequency for $\delta=0.2$ and $W = 4$, just above the crossover into the AMI-0
region (when coming from smaller values of $W$, for fixed $\delta$). 
The average DOS follows the bare distribution of $\varepsilon_i$ and it is simply given,
in the low temperature limit and for $W>U$, by the superposition of two rectangles.
In this regime in fact, the DMFT-TMT theory reduces to the
  superposition of isolated impurities, as we explained in Sec.~\ref{sec:model} [see eqs. (5-7)],
  and analytical expression can be derived.
    
The occupation per site and per spin as a function of on-site energy
corresponding to the DOS in Fig.~\ref{fig:VAMI_1} is shown in inset. Note that empty sites start to appear, 
besides those which are doubly- and singly-occupied. We have analyzed 
other values of $\delta$ for $U=3$, as well as results for $U=1$ and $U=2$,
and concluded that the AMI-0 arises for all $U$, as presented in
Fig.~\ref{fig:WxdopingHU}, arising directly from the metal for small $U$.

\begin{figure}
  \includegraphics[scale=0.75]{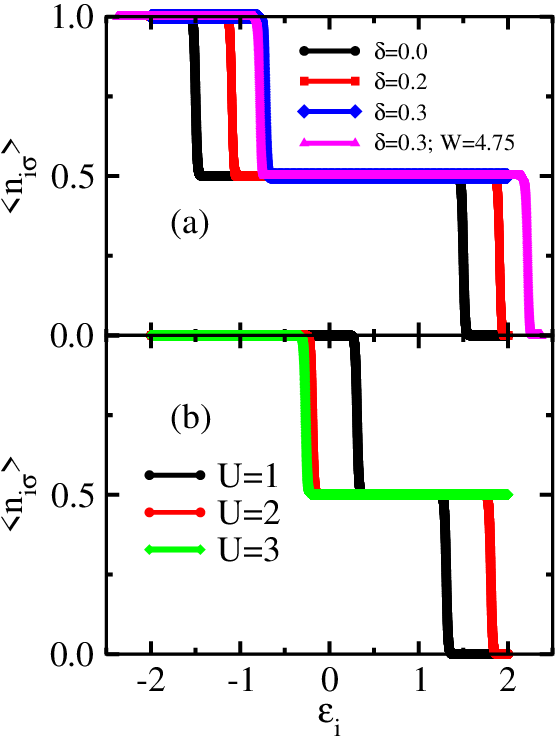}
  \caption{(a) Site occupation per spin as a function of the on-site
    energy for different values of doping and $W = 4$, except for the
    magenta curve with triangles, for which $W = 4.75$. $U=3$ and $T=0.01$.
    (b) Site occupation per spin as a function of the on-site energy
    for different values of the electronic correlation. $W=4$, $\delta=0.4$, and $T=0.01$.
    %
\label{nvsepsdifdelta} } 
\end{figure}

We shall now analyze how the AMI-0 region depends on doping $\delta$ and interaction $U$.
As doping increases, carriers are added to the system and more sites become doubly-occupied in comparison with a smaller $\delta$, as can be seen in Fig.~\ref{nvsepsdifdelta}(a). As a consequence, more disorder has to be added to the system [compare the results for $W=4.0$ and $W=4.75$ of the same figure for $\delta=0.3$] to empty some sites. The AMI-0 thus appears for larger values of disorder when the doping $\delta$ increases, as we observe in the phase diagrams of Fig.~\ref{fig:WxdopingHU}.


We now show in Fig.~\ref{nvsepsdifdelta}(b) the occupation per site and per
spin for $W=4$, $\delta=0.4$, and the different values of $U$. Since
single occupation is a characteristic of Mott insulators, the plateau
at $\left< n_{i\sigma} \right>=0.5$ becomes larger as $U$ increases. For the values of $W$ and
$\delta$ considered in the figure, the empty sites present for $U=1$
and $U=2$ become occupied as we move to $U=3$.
Thus, empty sites ``disappear'' with either the increase of doping
or correlations and, for large $U$, most of the disorder versus doping
phase diagram corresponds to an AMI without empty sites, as we describe
in detail in the following section.






\section{\label{sec:results} AMI phase results for $U=3.0$}

We shall now study the AMI without empty sites that appears by
increasing interaction $U$, sandwiched between the AMI-0
and the metallic phase on a large part of the phase diagram.
To this purpose, we consider the $U=3$ case (Fig. \ref{fig:WxdopingHU}c), and study the Anderson-Mott transition from the
disordered metal by increasing the disorder strength $W$ and various dopings $\delta$. We analyze in particular the behavior of typical and average DOS.\\

\subsection{Metal-insulator transitions} \label{mitU3}

\begin{figure}
	\includegraphics[scale=0.75]{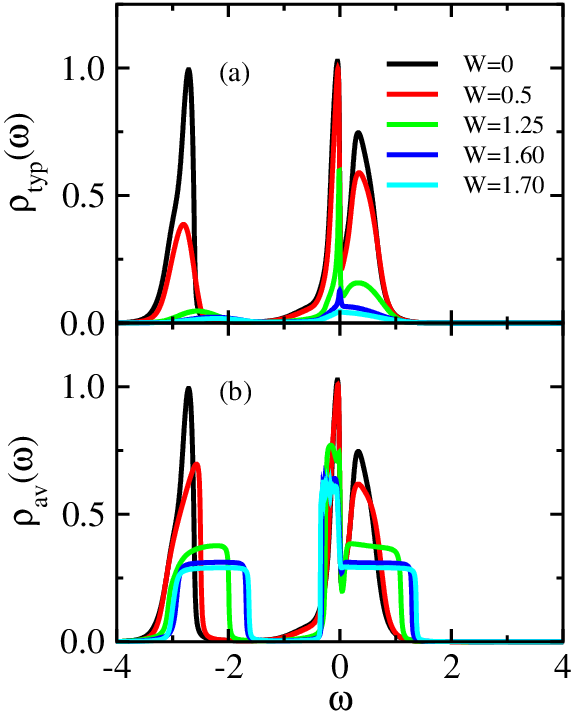}
	\caption{\label{fig:DOStypMetal} (a) Typical and (b) average DOS as a function of
	  frequency for different values of disorder $W$ and fixed doping $\delta=0.2$.
          $U=3$ and $T=0.01$.}
\end{figure}

The DMFT-TMT results for $\rho_{typ}(\omega)$ and $\rho_{av}(\omega)$ are shown,
respectively, in panels (a) and (b) of 
Fig.~\ref{fig:DOStypMetal} for fixed $\delta=0.2$ and different values of disorder
$W$. Since we have small doping, the typical DOS presents a structure of three peaks:
two Hubbard bands separated by an energy of the order of $U$ and a quasiparticle-like
  peak at the Fermi level $\omega=0$, as it has been previously reported in the clean case.\cite{Kajueter}
According to our results, this holds for small disorder 
as well, characterizing the system as a correlated metal in this region of parameters. 

As disorder $W$ increases, Anderson localization starts to play a role: its
effects can be seen by comparing the results for $\rho_{typ}(\omega)$ [panel (a)]
and $\rho_{av}(\omega)$ [panel (b)], since the former takes into account only
extended states, while the latter includes both extended and localized states of
the system. As $W$ increases, states at the band edges localize\cite{ziman} 
and we observe that the bands in the typical DOS become smaller. For even more disordered systems, $\rho_{typ}(\omega)$ vanishes on the whole frequency axis, signalizing that the system has gone
through a disorder-driven MIT. 
We notice that disorder acts differently on different energy scales. In $\rho_{typ}(\omega)$
Hubbard-like bands around larger values of energy shrink faster than the
quasiparticle-like one close to the Fermi level. 
Notice that  $\rho_{av}(\omega)$ remains instead finite at the Anderson-Mott
  transition. The general effect of disorder appears to be a spreading in energy
  of the spectral weight, both in the Hubbard bands and at the low-energy quasi-particle
  peak. This spreading is not symmetrical like at half-filling (see the results in
    Appendix \ref{half-filling}), because of the combined effect of disorder and doping.

\begin{figure}
  \includegraphics[scale=0.75]{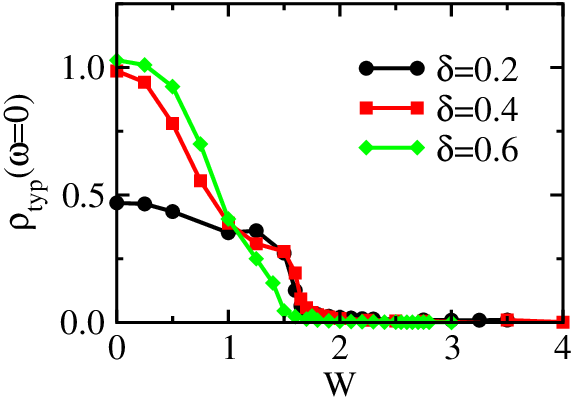}
\caption{\label{fig:DOStypdifdelta} Typical DOS at the Fermi level as a function of
  disorder for different values of doping. Results for $\delta=0.2$ correspond to
  $\rho_{typ}(\omega)$ shown in panel (a) of Fig.~\ref{fig:DOStypMetal}. $U=3$ and $T=0.01$.}
\end{figure}

To determine the critical disorder $W_c$ at which the MIT takes place, 
it is easier to track the typical DOS at the Fermi level as a function of disorder, 
as we display in Fig.~\ref{fig:DOStypdifdelta} for different values of doping. 
This quantity plays the role of an order parameter for the disorder induced MIT, 
since it is different from zero in the metallic region ($W<W_c$) and is zero in 
the Anderson-Mott insulator ($W>W_c$). Based on the behavior
of $\rho_{typ}(\omega=0)$ as a function of disorder, we have 
determined the transition line between the metallic and AMI phases shown in 
Fig.~\ref{fig:WxdopingHU} (red filled dot line). Notice that for $\delta=0.2$ the maximum of 
$\rho_{typ}(\omega)$ is close to $\omega=0$, but not exactly at $\omega=0$ (in 
accordance with the results of Ref.~\onlinecite{Kajueter}); this explains why 
$\rho_{typ}(\omega=0) \approx 0.5$ for the clean system in
Fig.~\ref{fig:DOStypdifdelta}, instead of the maximum value of $\approx 1$ for
$\rho_{typ}(\omega)$ seen in Fig.~\ref{fig:DOStypMetal}(a).


A key observation is that
there exists only a small dependence of $W_c$ with 
doping $\delta$. As mentioned in Sec.~\ref{sec:PhaseDiagram}
and observed in Fig.~\ref{fig:DOStypMetal}(a),
the transition to the AMI takes place when all extended states of the system localize. 
For fixed $U=3$
(look now at Fig.~\ref{compU1U3}), the range in energy
where $\rho_{typ}(\omega)$ extends is roughly doping independent.
If doping increases, we observe mainly a transfer
of spectral weight from above the Fermi level to energies 
below it. This might justify the fact that $W_c$ is practically constant for small 
to moderate $\delta$. 
A similar behavior is observed for $U=1$, but not in the
intermediate case of $U=2$, as we have discussed in Sec.~\ref{sec:PhaseDiagram}.

More interesting is the fact that $W_c$ for the doped case is smaller
than $U$, which is the critical disorder at which the transition from the Mott 
insulator to AMI-0 is seen at half-filling (end of orange line in 
Fig.~\ref{fig:WxdopingHU}). This means that the doped Mott insulator
is more susceptible to disorder induced localization, as we have already
mentioned in Sec.~\ref{sec:PhaseDiagram}. By
introducing carriers into the system in fact, a narrow band rises within the gap, as
seen is our results in Fig.~\ref{fig:DOStypMetal}(a). By adding disorder to the
doped system, this narrow band localizes at a disorder strength which is 
smaller than
the one required to Anderson localize the Mott insulator, which requires that the wide Mott gap is filled due to disorder
effects. (For details on how the transition is approached at 
half-filling see Appendix \ref{half-filling}.)
As mentioned in the previous section, this is not observed for $U=1$ and $U=2$,
probably because for these values of $U$ the system is in a metallic
state at half-filling, and the wide Mott gap is replaced by a wide band
  of itinerant states around the Fermi level, that can Anderson localize only
at higher disorder strengths.

\begin{figure}
\includegraphics[scale=0.75]{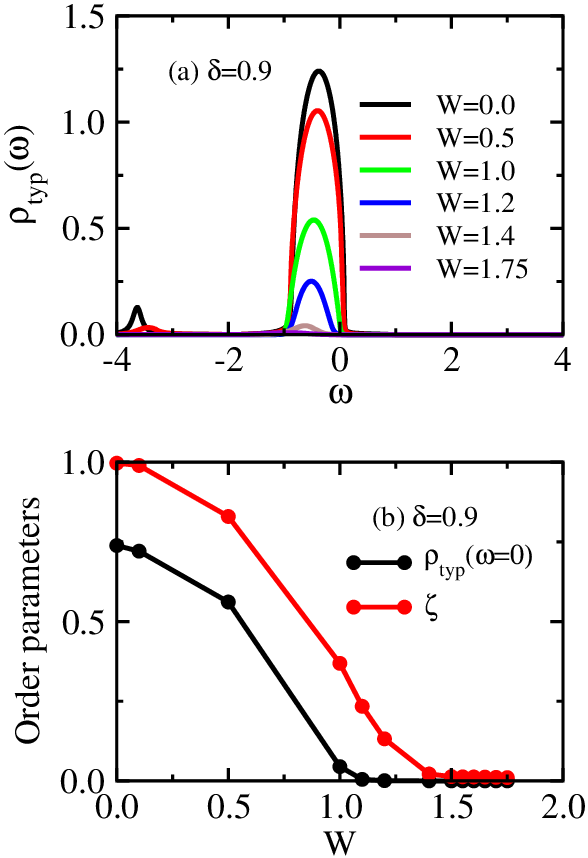}
\caption{\label{bandinsulator} (a) Typical DOS as a function of energy for
different values of disorder $W$ and fixed doping $\delta=0.9$. (b) Typical DOS 
at the Fermi level and $\upzeta=\int_{-\infty} ^{\infty} \rho_{typ}(\omega)d\omega$ 
as a function of disorder corresponding to the results in (a). $U=3$ and $T=0.01$. }
\end{figure}

For large doping ($0.75 < \delta < 1.0$), we observe a region within 
the Anderson-Mott insulator that, with abuse of language, we identify as a band AMI.
Starting with the clean system,
$\rho_{typ}(\omega)$ shrinks as disorder increases, similar to what happens
for small doping. However, 
for large doping, $\rho_{typ}(\omega=0)$ becomes zero for a smaller
value of disorder ($W_{c}$) than that at which the whole band vanishes ($W_{c2}$).
This means that, differently from the low doping case where the whole $\rho_{typ}(\omega)$
  vanishes, for $W_c < W < W_{c2}$ the system has still a band of extended
states, which is located below the Fermi energy. This behavior is exemplified
in Fig.~\ref{bandinsulator} for the case of $\delta=0.9$: panel (a) shows
$\rho_{typ}(\omega)$ for different values of disorder, while panel (b)
presents both $\rho_{typ}(\omega=0)$ and 
$\upzeta=\int_{-\infty} ^{\infty} \rho_{typ}(\omega)d\omega$ as a 
function of $W$.
As we can see, for this value of doping, the system enters into the insulating
phase [$\rho_{typ}(\omega=0)$=0] at $W_{c} \approx 1.1$, while all states are
localized ($\upzeta=0$) only at $W_{c2} \approx 1.75$.
This behavior of $\rho_{typ}(\omega=0)$ at $W_{c}$ is reminiscent of that of
  a band insulator, though the total spectral intensity $\rho_{av}(\omega=0)$
  remains in all cases finite at the Fermi level. Note that $W_{c2}$ 
approximately coincides with the disorder at which the system enters the AMI 
region for $\delta < 0.6$, as expected if the vanishing of $\upzeta$
mainly depends on the $U$ value.


\subsection{Character of the Anderson-Mott insulator}

\begin{figure}
\includegraphics[scale=0.75]{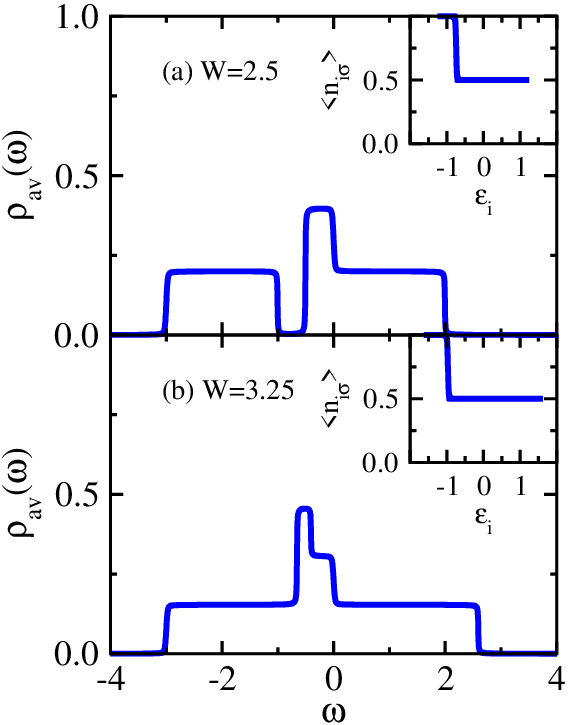}
\caption{\label{fig:DOSave} Average DOS as a function of energy for $U=3$ and 
(a) $W=2.50$ and (b) $W=3.25$. The insets show the occupation number per site
and per spin as 
a function of the on-site energy for the same parameters of the results in the main 
panels. $\delta=0.2$ and $T=0.01$.}
\end{figure}

We want now to characterize the physical properties of the AMI region.
  To this purpose, we shall focus on 
the (arithmetic) average DOS $\rho_{av}(\omega)$ [defined in eq.~(\ref{eq:rhoAve})],
which can be directly connected to spectroscopic experiments.
Fig.~\ref{fig:DOSave} shows two examples of $\rho_{av}(\omega)$ for $\delta=0.2$
and $U= 3$:
one for which the disorder $W<U$ [panel (a)] and another for which $W>U$ [panel (b)].
In both cases the system is in the AMI region, where all states are localized. Since 
the typical DOS is zero, meaning that there is no bath for
the impurities to hybridize, within our TMT approximation impurity sites are  effectively in the atomic
limit, and the DOS can be calculated with eqs. (5-7) described in Sec. \ref{sec:model}.
In this case, in the absence of disorder $\varepsilon_i$
the DOS of a single-impurity problem presents two Dirac delta peaks, one at $\omega=-U/2$
and another at $\omega=U/2$. As disorder $\varepsilon_i$ is added, these peaks
spread in energy following the flat uniform distribution of the disorder.
If disorder is large enough, these rectangles overlap at small frequencies, giving rise
to the form of the average DOS seen in Fig.~\ref{fig:DOSave}.

By looking at the results in panel (a), for $W<U$, we observe that  
there is a well defined  gap at negative energies in $\rho_{av}(\omega)$. This profile for the DOS reminds us 
that of the slightly doped Mott insulator. In the case of $W>U$ [panel (b)], 
on the contrary, a gap is not seen in $\rho_{av}(\omega)$ anymore. This reminds
us of an Anderson insulator.
In the insets of the panels, we show the
corresponding occupation $\left< n_{i\sigma} \right>$ per site and per spin as a function of the
on-site energy $\varepsilon_i$. In both cases,\cite{Carol_twofluids} there are
sites that are doubly occupied ($\left< n_{i\sigma} \right> =1$), as in an Anderson insulator, and sites 
that are singly-occupied ($\left< n_{i\sigma}\right> =0.5$), as in a Mott insulator (but no
empty site as in the AMI-0 region). 
Since the two systems have characteristics of Anderson and Mott insulators,
we identify both cases as AMI in the phase diagram of Fig.~\ref{fig:WxdopingHU}. 
However, we are tempted to say that in the case of panel (a) the role played by Mott 
mechanism of localization is stronger than that of Anderson effects. Indeed,
the correlation $U$ is larger than disorder $W$ and a gap is observed in the 
average DOS. On the other hand, Anderson mechanism may dominate over the Mott 
one in the case where disorder is larger than correlations [panel (b)].
In the phase diagram, the first case is observed whenever $W_c<W<U=3$, while 
the second corresponds to $W>U=3$. Similar behavior occurs for $U=2$ and $\delta>0.5$,
for which there is a region where $W_c<U$.

We recall that at half-filling ($\delta=0$) the Mott dominated region
  for $W< W_c\approx U$
  presents only singly $\left< n_{i\sigma} \right>=0.5$ occupied sites, being double occupation
  $\left< n_{i \sigma} \right> =1$ absent. \cite{Carol2}
%
On the other hand, for $W>U$, there exist doubly- and singly-occupied
sites, as well as empty ones, and the average DOS has no gap. The 
presence of this third kind of sites - the empty ones - gives rise to the
AMI-0 region, as we described in subsection \ref{Vshaped}.

In the next subsection, we discuss how the doped system crossovers
from the AMI region to the AMI-0 one.


\subsection{Crossover to the AMI-0 region}

\begin{figure}
  \includegraphics[scale=0.75]{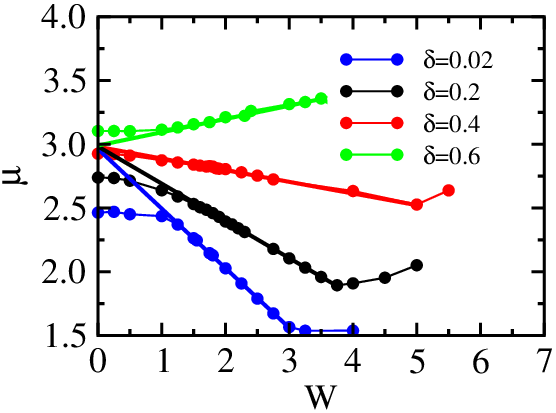}
  \caption{\label{muvsW} Chemical potential $\mu$ obtained to
    keep the values of $\delta$ fixed as a function of disorder strength.
    Straight lines correspond to fittings of the numerical results
  in the range where a linear behavior is observed; note that the lines
  meet at the point $\mu=U=3$. $U=3$ and $T=0.01$.}
\end{figure}

Finally, it is useful to see the behavior of the chemical potential $\mu$ as
a function of disorder for fixed doping, showed in Fig.~\ref{muvsW}, again for $U=3$.
In the
metallic phase (for small values of $W$), the $\mu$ vs. $W$ curve is markedly 
horizontal. By entering into the AMI, above $W \approx 1.75$, the $\mu$ vs. $W$
follows closely a linear law $\mu=(\delta-0.5)W+U$. By further increasing disorder,
the $\mu$ vs. $W$ curve displays once again a sharp change in slope when entering into the
AMI-0 that presents empty sites. For $\delta$ values in this AMI-0 region,
$\mu$ always increases with $W$.

The linear $\mu$ vs. $W$ behavior in the AMI region allows us to establish an
equation determining the disorder dependence of the line separating the
AMI-0 and AMI regions.
We profit again that within the DMFT-TMT method in the AMI region each impurity site is in the atomic
limit (as described in Sec. \ref{sec:model}). In this case, the site is empty
if its on-site energy $\varepsilon_d = \varepsilon_i-\mu> 0$ (eq. 6).
The value of the disorder where the first empty site forms must coincide with
the highest possible value of on-site energy $\varepsilon_i = W / 2$, i.e.
$W / 2 = \mu$. Plugging this value of $\mu$ into the $\mu$ vs. $W$ linear
relation, which holds up to the crossover to the AMI-0 region, we obtain:
\begin{equation}
W=\frac{U}{1-\delta}. \label{EquationV}
\end{equation}
Equation \ref{EquationV}, which we display in Fig.~\ref{fig:WxdopingHU} as a
dashed gray line, well describes the crossover between AMI and AMI-0 regions that we establish numerically (brown diamond symbols in
the figure).

    
\section{Conclusion} \label{sec:conclusion}

In this work, we solved the Anderson-Hubbard model in the doped case by using the combination of Dynamical Mean Field Theory and Typical Medium Theory. The former describes the Mott transition, while the latter takes into account Anderson localization effects. We built the disorder versus doping phase diagram for three values of $U$: $U=1$, $U=2$, and $U=3$, in units of the clean, non-interacting bandwidth. 
For any interaction, there is a region of the phase diagram where we observe an Anderson-Mott insulator similar to the one that exists at half-filling, with the presence of empty sites in the system (AMI-0). 
As doping (and thus the number of carriers in the system) increases, the empty sites
that exist at small doping become occupied, giving rise to a different Anderson-Mott insulator (AMI).
When the electronic interaction becomes stronger, 
the AMI wedges between the metallic phase and the AMI-0,
occupying large part of the phase diagram in the strongly correlated regime, for $U=3$.
    This is a consequence of the fact that the disorder-driven MIT takes place for a much smaller
    disorder strength in the doped, strongly correlated regime than in the weakly correlated
    regime or
    at half-filling.
An intermediate behavior should appear in the intermediate correlated regime,
with the critical disorder $W_c$ monotonically decreasing with doping, as portrayed
in the phase diagram (Fig.~\ref{fig:WxdopingHU}) for $U=2$.

Upon doping, the properties of the system are therefore strongly determined by the combined effect of disorder,
interaction, and doping to form an insulator presenting at the same time Anderson and Mott-Hubbard
features.
The evolution of the phase diagram as a function of disorder and doping, that we presented in this work from the weak
  to strongly correlated regime, should determine the universal properties of the disorder-driven MIT.

\begin{acknowledgments}
  We acknowledge Eduardo Miranda and Marcelo Rozenberg for discussions
  and Jaksa Vu\v{c}i\v{c}evi\'c and Darko Tanaskovi\'c for the development
  of the IPT single-impurity code used in our calculations.
  This work is supported by FAPEMIG, CNPq (in particular through
  INCT-IQ 465469/2014-0), and CAPES (in particular through
  programs CAPES-COFECUB-0899/2018 and CAPES-PrInt-UFMG (M.C.O.A.)).  M.C. acknowledges support from the ANR grant NEPTUN no. ANR-19-CE30-0019-04.
\end{acknowledgments}

\appendix

\section{Spectra and metal-insulator transition at strong $U>3$} \label{largeU}

\begin{figure}
	\includegraphics[scale=0.75]{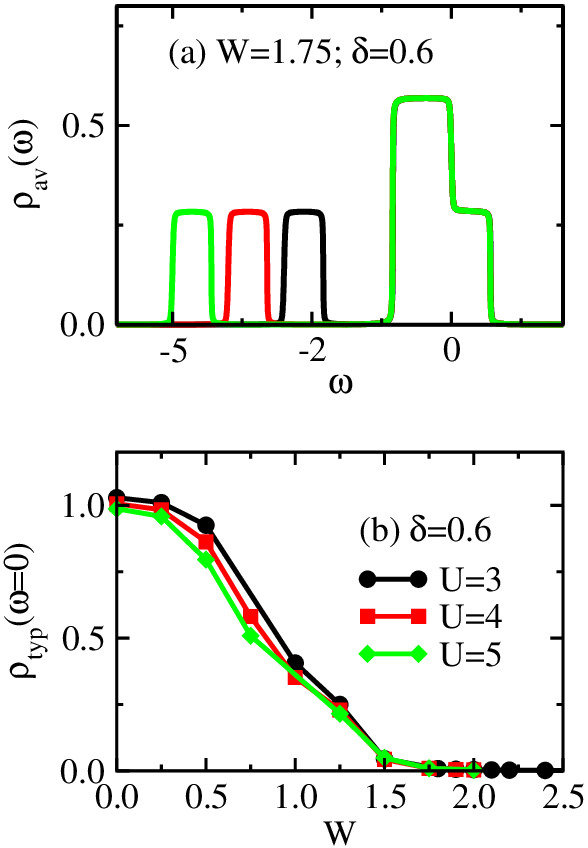}
	\caption{Results for large values of the interaction, $U=3$, $U=4$, and $U=5$. (a) Average
	  DOS as a function of frequency when $W=1.75$.
          (b)~Typical DOS at the Fermi level as a function
		of $W$, showing that the order parameter goes to zero for the same
		value of $W_c$ (for $U \geq 3$). $\delta=0.6$ and $T = 0.01$.
	} \label{HightU}
\end{figure}

We shall show here that the results on the MIT that we established for the $U=3$ case are qualitatively similar for larger interaction $U$. Interaction has the effect of changing the position of the DOS bands. In the presence of doping, only the low energy band moves proportionally to the $U$ value. A quasiparticle-like band remains around the Fermi level; its position does not change by increasing $U$ to keep the doping $\delta$ fixed. In Fig.~\ref{HightU}(a), where we display the
  average DOS $\rho_{av}(\omega)$, we can see an example of this behavior: increasing the interaction does not change the Fermi level band. 
      This causes the value of $W_c$ not to change, as can be seen in Fig.~\ref{HightU}(b). This phenomenon only occurs when the bands are far apart
    (in the Mott regime $W<U$ of the doped AHM).

\section{Results for $U = 3$ at half-filling} \label{half-filling}

\begin{figure}
	\includegraphics[scale=0.75]{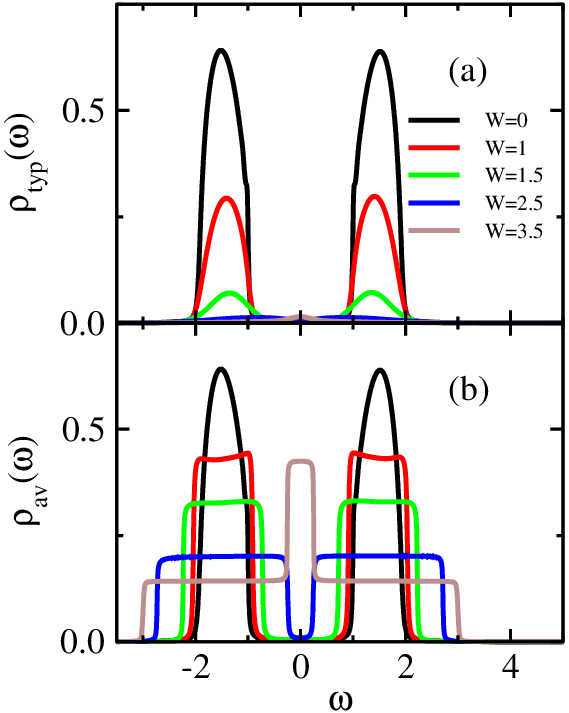}
	\caption{\label{DOShalf-filling} (a) Typical and (b) average DOS as a function of
		energy for different values of disorder $W$ at half-filling.  $U=3$ and $T=0.01$.
	}
\end{figure}

The evolution of the typical and average DOS as disorder increases for the 
case of $U=3$ and no doping is presented in Fig. \ref{DOShalf-filling}. The 
clean system has a Mott gap, which starts to be filled with localized states 
as disorder increases. The gap eventually closes and the system becomes
  an Anderson-Mott insulator, as illustrated in Fig.~\ref{DOShalf-filling} for
  $W = 3.5$. At half-filling, the Anderson-Mott insulator always presents
empty sites and thus corresponds to the AMI-0 defined in the main text. 
This behavior is different than the one presented in Fig. \ref{fig:DOStypMetal} for the doped case, where the clean system is a metal.


\bibliographystyle{apsrev4-1}

\end{document}